\documentclass[letterpaper]{jpconf}
\usepackage{graphicx}
\begin{document}
\title{Proton-neutron pairing correlations in the nuclear shell model}

\author{Lei Yang and Yu-Min Zhao}

\address{Department of Physics, Shanghai Jiao Tong University, Shanghai, 200240, China}

\author{S. Pittel and B. Thakur}

\address{Bartol Research Institute and Department of Physics and
Astronomy, University of Delaware, Newark, Delaware 19716, USA}

\author{N. Sandulescu}

\address{Institute of Physics and Nuclear Engineering, 76900
Bucharest, Romania}

\author{A. Poves}

\address{Departamento de Fisica Teorica and IFT UAM/CSIC, Universidad Autonoma de Madrid,
    28049, Madrid Spain.}

\ead{pittel@bartol.udel.edu}

\begin{abstract}
A shell-model study of proton-neutron pairing in $f-p$ shell nuclei
using a parametrized hamiltonian that includes deformation and
spin-orbit effects as well as isoscalar and isovector pairing is
reported. By working in a shell-model framework we are able to
assess the role of the various modes of proton-neutron pairing in
the presence of nuclear deformation without violating symmetries.
Results are presented for $^{44}Ti$, $^{46}Ti$ and $^{48}Cr$.
\end{abstract}

\section{Introduction}

It is generally believed that $proton-neutron$ ($pn$) pairing is
important in nuclei with roughly equal numbers of neutrons and
protons \cite{Goodman}. The standard technique for treating these
correlations is through BCS or HFB approximation, generalized to
include the $pn$ pairing field in addition to the $nn$ and $pp$
pairing fields. Questions arise, however, as to whether these
methods can adequately represent the physics of the competing modes
of pair correlations, without full restoration of symmetries
\cite{DP}.

Important insight into this issue has been achieved recently in the
context of exactly-solvable models that include these different
pairing modes. Analysis of the SO(8) model, in which isoscalar and
isovector pairing act in either a single active orbital or a series
of degenerate orbitals, suggests that isospin restoration or
equivalently quartet correlations are extremely important,
especially near $N=Z$ \cite{DP}. More recent studies, carried out
for models involving non-degenerate orbitals \cite{Errea}, reinforce
earlier conclusions as to where isoscalar pairing correlations
should be most important \cite{Satula},\cite{Sandulescu}.
Furthermore, they make possible the description of deformation, as
is critical for systems with $N \approx Z$,  by treating the
non-degenerate orbitals as Nilsson-like. However, it is still not
possible in these models to restore symmetries, either rotational or
isospin.

As a consequence, there still remain many open issues concerning the
role of the different possible modes of pairing in $N \approx Z$
nuclei. In this work, we report a systematic study of pairing
correlations in the context of the nuclear shell model, whereby
deformation can be readily included and symmetries maintained. In
this way, we are able to address many of the open issues on the role
of the various pairing modes in the presence of nuclear deformation.

An outline of the paper is as follows. In Section 2 we describe our
model and in Section 3, we report some of the key results we have
obtained, which are then summarized in Section 4.

\section{Our model}

To address in a systematic way the role of pairing correlations in
the presence of nuclear deformation, we consider neutrons and
protons restricted to the orbitals of the $1f-2p$ shell outside a
doubly-magic $^{40}Ca$ core and interacting via a schematic
hamiltonian

\begin{equation}
H= \chi \left( Q \cdot Q + a P^{\dagger} \cdot P + b S^{\dagger}
\cdot S + \alpha \sum_i \vec{l}_i \cdot \vec{s}_i \right)
\label{hamiltonian}
\end{equation}
where $ Q = Q_n +Q_p$ is the mass quadrupole operator, $P^{\dagger}$
creates a correlated $L=0$, $S=1$, $J=1$, $T=0$ pair and
$S^{\dagger}$ creates an $L=0$, $S=0$, $J=0$, $T=1$ pair. The first
term in the hamiltonian produces rotational collective motion,
whereas the second and third term are the isoscalar and iosvector
pairing interactions, respectively. The last term is the one-body
part of the spin-orbit interaction, which splits the $j=l \pm 1/2$
levels with a given $l$.

We carry out calculations systematically as a function of the
various strength parameters. We begin with pure SU(3) rotational
motion associated with the $Q \cdot Q$ interaction and then
gradually ramp up the various SU(3)-breaking terms to assess how
they affect the rotational properties. This includes the isocalar
and isovector pairing interactions and the spin-orbit term.

We first consider the nucleus $^{44}Ti$, with $N=Z=2$, and then
systematically increase $N$ and $Z$ to study the role of the number
of active neutrons and protons, e.g. whether there is an excess of
one over the other and whether the nucleus is even-even or odd-mass.
The nuclei we have treated are $^{44}Ti$ ($N=Z=2$), $^{45}Ti$
($N=2$, $Z=3$), $^{46}Ti$   ($N=2$, $Z=4$), $^{46}V$ ($N=3$, $Z=3$),
and $^{48}Cr$ ($N=4$, $Z=4$). Some of the observables we have
studied are (1) the energies and associated BE(2) values of the
lowest rotational band, (2) the number of $J^{\pi}=6^+$ $T=1$ pairs,
(3) the number of $J^{\pi}=0^+$, $T=1$ ($S$) pairs and  (4) the
number of $J^{\pi}=1^+$, $T=0$ ($P$) pairs. In the following
section, we present selected results for $^{44}Ti$, $^{46}Ti$ and
$^{48}Cr$.

\section{Calculations}
\subsection{{\it Optimal} hamiltonian}

We first ask whether the hamiltonian (\ref{hamiltonian}) has
sufficient flexibility to describe the nuclei under investigation.
Without making an effort towards an absolute fit, we find that the
choice $ \chi = -0.05 ~MeV,~a=b=12$, and $\alpha= 20$ gives an
acceptable fit to the spectra of all the nuclei we have considered.
This is illustrated in figure 1 for $^{44}Ti$, $^{46}Ti$ and
$^{48}Cr$. As can be seen, the well-known non-rotational character
of $^{44}Ti$ is reproduced by our calculations, as are the highly
rotational patterns seen experimentally for the heavier nuclei. As
we will see later, even the experimentally observed backbend in
$^{48}Cr$ is acceptably reproduced with this hamiltonian. We refer
to the choice $a=b$ in the {\it optimal} hamiltonian as the SU(4)
choice, from the dynamical symmetry that derives from this choice of
parameters in the SO(8) model.

\begin{figure}
\begin{center}
\includegraphics[width=12cm]{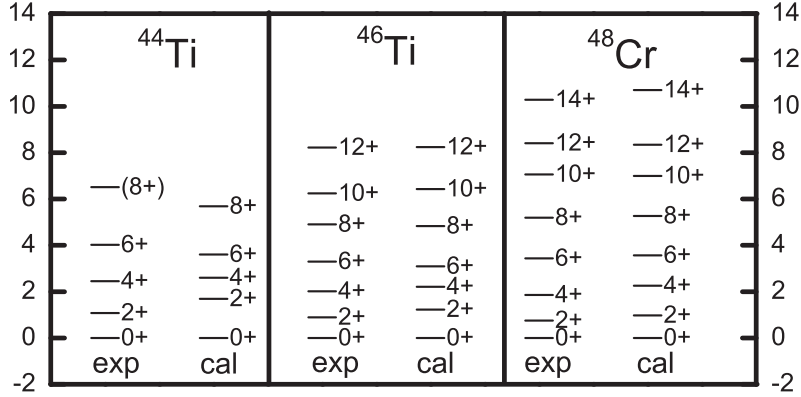}
\end{center}
\vspace{-1cm} \caption{\label{fig1}Comparison of experimental
spectra for $^{44}Ti$, $^{46}Ti$ and $^{48}Cr$ with the calculated
spectra obtained using the {\it optimal} hamiltonian described in
the text. All energies are in $MeV$.}
\end{figure}

\subsection{$^{44}Ti$}

We next turn to the nucleus $^{44}Ti$, with two active neutrons and
two active protons. In figure 2, we show the calculated energy
splittings $E_{I}-E_{I-2}$ associated with the ground-state band as
a function of the strength parameters $a$ and $b$ that define the
isoscalar and isovector pairing interactions, respectively. For
these calculations we assumed a quadrupole strength of
$\chi=-0.05~MeV$ and no spin-orbit interaction. What we see is that
the isoscalar and isovector pairing interactions have precisely the
same effect on the properties of the ground state rotational band,
{\it in the absence of any spin-orbit interaction}. Precisely the
same conclusion derives when we consider the effect of isoscalar and
isovector pairing on other observable properties.

\begin{figure}[h]
\begin{center}
\hspace{5cm}
\includegraphics[width=20cm]{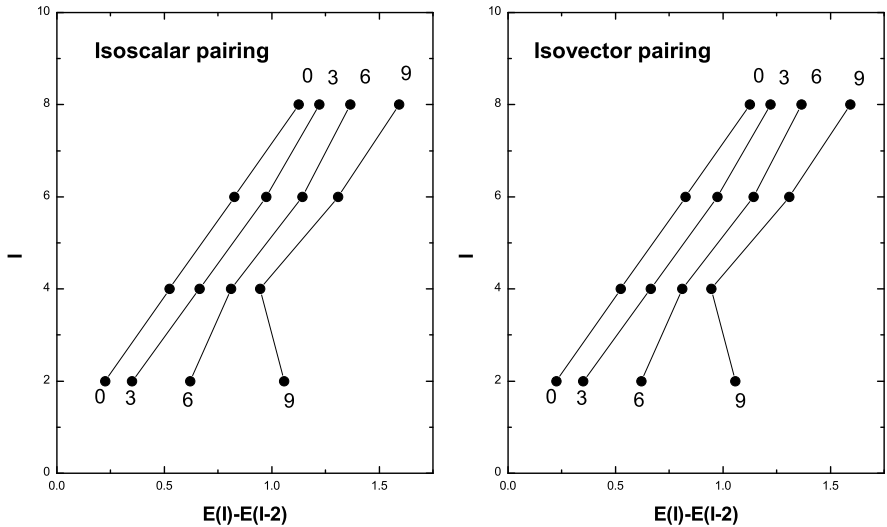}
\end{center}
\vspace{-7.5cm} \caption{\label{fig2} Spectra of the ground state
band of $^{44}Ti$ as a function of the strength of the isoscalar
pairing interaction (left panel) and of the isovector pairing
interaction (right panel), in each case with no spin-orbit term
present. }
\end{figure}

Some other results from our study of $^{44}Ti$ are illustrated in
figure 3. In panel (a), we show the spectrum that derives solely
from turning on a strong spin-orbit force. We see that the spectrum
is still highly rotational, despite the fact that the resulting
single-particle energies are no longer SU(3)-like. To obtain the
physical spectrum with a non-rotational character, it is thus
essential to have pairing. It is usually accepted that it is the
non-SU(3) order of the single-particle levels that is responsible
for the non-rotational character seen \cite{Bhatt}, a conclusion not
supported by our results. In panel (b), we address the issue of how
isoscalar pairing is affected by inclusion of a spin-orbit force. We
see that the number of isoscalar pairs -- defined as the expectation
value of $P^{\dagger}\cdot P$ -- decreases rapidly with an
increasing spin-orbit interaction, especially for the lowest angular
momentum states of the ground band. The mechanism whereby the
spin-orbit interaction suppresses isoscalar pairing has been
discussed recently in ref. \cite{Bertsch}.

\begin{figure}
\begin{center}
\includegraphics[width=16cm]{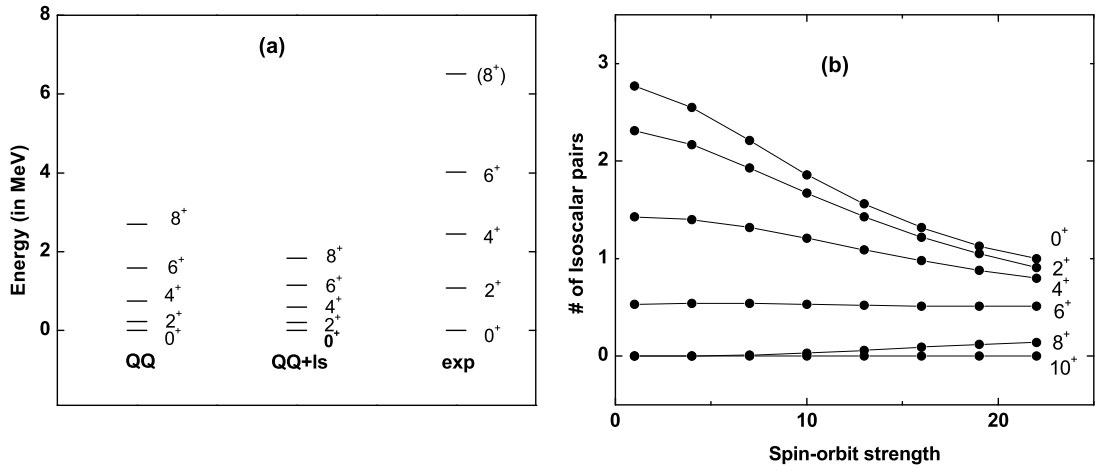}
\end{center}
\vspace{-2cm} \caption{\label{fig3} Results from calculations
carried out for $^{44}Ti$. (a) Comparison of he experimental
spectrum with spectra obtained with a pure $Q \cdot Q$ interaction
and with both a $Q \cdot Q$ interaction and a spin-orbit term. (b)
Number of isoscalar pairs as a function of the spin-orbit strength,
for the various states in the ground band. }
\end{figure}

\subsection{$^{46}Ti$}

Next we turn to $^{46}Ti$ with two additional neutrons present. Here
too we compare the effect of the isoscalar and isovector pairing
interactions on deformation, showing the results in figure 4 with no
spin-orbit term present. Here the effect of isoscalar pairing is
strongly suppressed relative to isovector pairing, suggesting that
even without a spin-orbit term isoscalar pairing is very strongly
focused on those nuclei with $N=Z$ with a slight excess being
sufficient to suppress this pairing mode.

\begin{figure}
\begin{center}
\includegraphics[width=20cm]{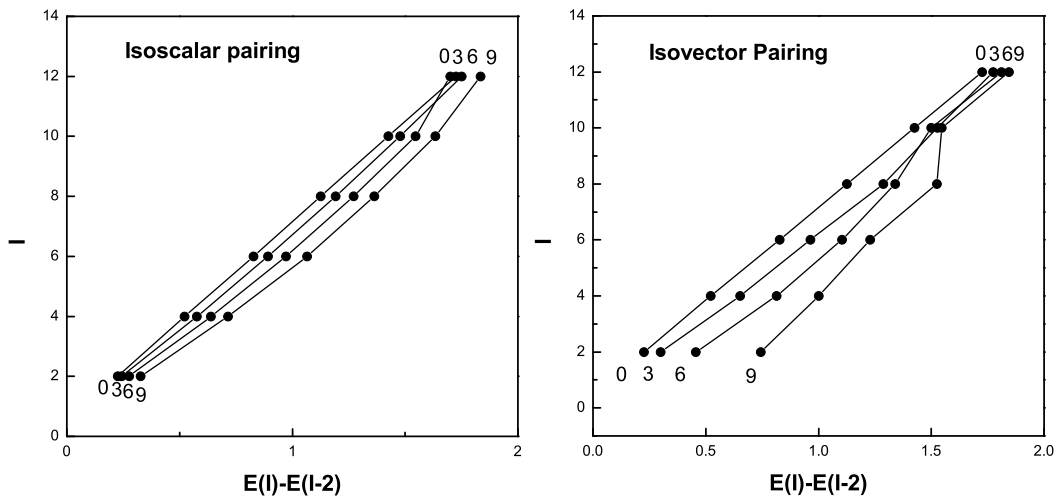}
\end{center}
\vspace{-6cm} \caption{\label{label}Spectra of the ground state band
of $^{46}Ti$ as a function of the strength of the isoscalar pairing
interaction (left panel) and of the isovector pairing interaction
(right panel), in each case with no spin-orbit term present.}
\end{figure}

\subsection{$^{48}Cr$}

Lastly, we turn to $^{48}Cr$, which again has $N=Z$, but now with
two quartet-like structures present. Here we assume as our starting
point both the optimal quadrupole-quadrupole force and one-body
spin-orbit force and then ramp up the two pairing strengths from
zero to their optimal values. The results are illustrated in figure
5, for scenarios in which we separately include isoscalar pairing,
isovector pairing and and SU(4) pairing with equal strengths.

As a reminder, the experimental spectrum for $^{48}Cr$ shows a
backbend near $I=12$, which as noted earlier is reproduced by our
{\it optimal} hamiltonian. The results of figure 5 make clear that
(a) the backbend cannot be reproduced with pure isoscalar pairing,
but requires isovector pairing as well and (b) there is no
significant difference between the results obtained with pure
isovector pairing and SU(4) pairing. The backbend in $^{48}Cr$ was
discussed extensively in the context of a shell-model study with a
fully realistic hamiltonian in \cite{Poves}, where it was first
shown to derive from isovector pairing. Our results are in agreement
with that earlier conclusion.
\begin{figure}[h]
\begin{center}
\includegraphics[width=20cm]{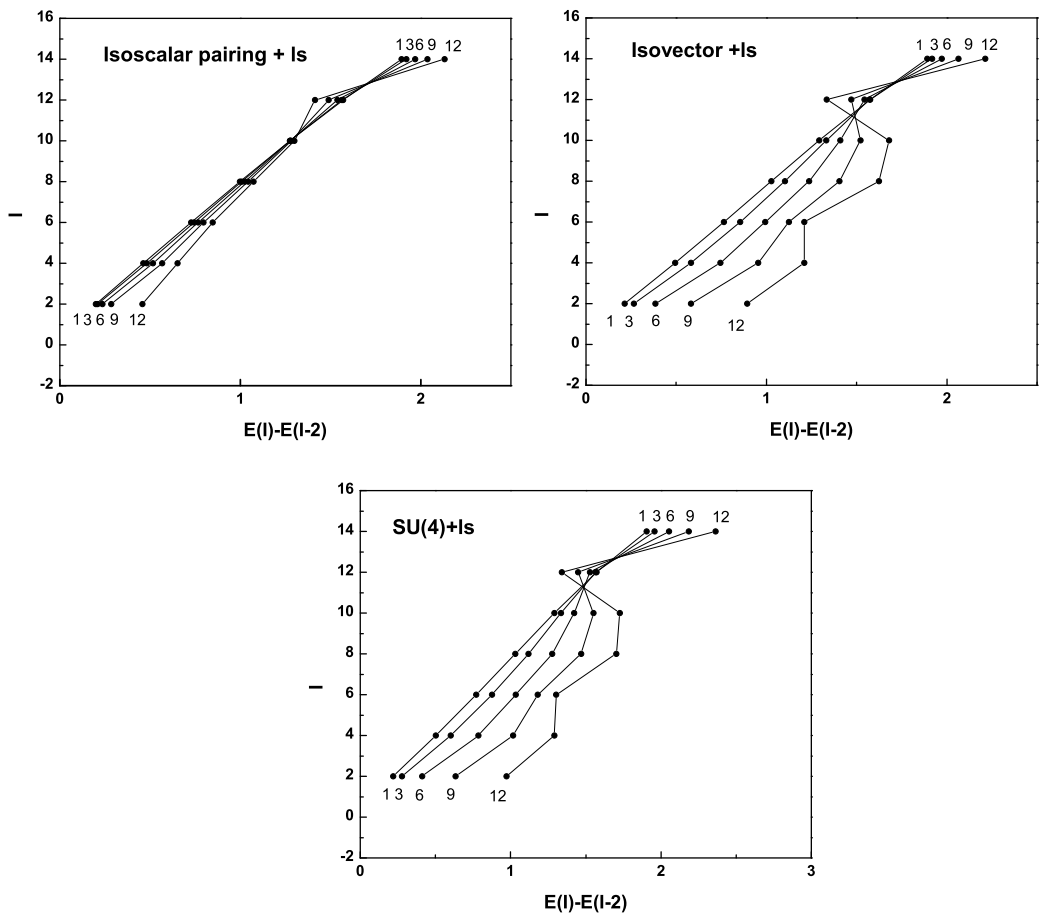}
\end{center}
\vspace{-1cm} \caption{\label{fig5}Calculated splittings in
$^{48}Cr$ ground band, for isovector, isoscalar, and SU(4) pairing,
respectively,  as described in the text.}
\end{figure}

\section{Summary and Concluding Remarks}

In this work, we have reported a shell-model study of proton-neutron
pairing in $f-p$ shell nuclei using a parametrized hamiltonian that
includes deformation and spin-orbit effects as well as both
isoscalar and isovector pairing. By working in a shell-model
framework we are able to assess the role of the various modes of
proton-neutron pairing in the presence of nuclear deformation
without violating symmetries.

We first showed that our parametrized hamiltonian has enough
flexibility to be able to provide a reasonable description of the
evolution of nuclear structure properties in this region. We then
probed the role of the various modes of pairing on deformation with
or without a spin-orbit term. We did this as a function of the
number of neutrons and protons, so as to assess the role both of a
neutron excess and of the number of active particles.

Some of the conclusions that emerged are: (1) in the absence of a
spin-orbit term, isoscalar and isovector pairing have identical
effects at  $N=Z=2$, but that isoscalar pairing ceases to have an
appreciable effect for nuclei with just two excess neutrons; (2) the
non-rotational character of $^{44}Ti$ cannot be explained solely in
terms of spin-orbit effects but requires pairing for its
understanding; (3) in the presence of a spin-orbit interaction,
isoscalar pairing is suppressed even at $N=Z$, and (4) the known
backbend in $^{48}Cr$ has its origin in isovector pairing.

\ack
This paper is dedicated by one of the authors (S.P.) to the
memory of Marcos Moshinsky, colleague, teacher and friend. The work
reported herein began while S.P. and N.S. were visiting the Consejo
Superior de Investigaciones Cient\'{i}ficas in Madrid, whose
hospitality is gratefully acknowledged. Much of it was carried out
while L.Y. was visiting the Bartol Research Institute of the
University of Delaware, whose hospitality is likewise acknowledged.
The work of S.P., B.T. and L.Y. was supported in part by the
National Science Foundation under grant \#s PHY-0553127 and
PHY-0854873, that of N.S. by the Romanian Ministry of
Education and Research through CNCSIS grant Idei nr. 270 and that of A.P. by the projects FPA2009-13377 MICINN(Spain)
and HEPHACOS S2009/ESP-1473 Comunidad de Madrid(Spain).

\end{document}